\documentclass[a4paper]{article}

\usepackage{spconf}
\usepackage{bm}
\usepackage{graphicx}
\usepackage{float}
\usepackage{ascmac} 
\usepackage{fancybox}
\usepackage{float}
\usepackage{tabularx}
\usepackage{booktabs}
\usepackage{amssymb}
\usepackage{multirow}
\usepackage{cite}
\usepackage{url}
\usepackage{setspace}
\usepackage{color}
\AtBeginDocument{
  \abovedisplayskip     =.8\abovedisplayskip
  \abovedisplayshortskip=.8\abovedisplayshortskip
  \belowdisplayskip     =.8\belowdisplayskip
  \belowdisplayshortskip=.8\belowdisplayshortskip}

\title{Multimodal Attention Fusion for Target Speaker Extraction}
\name{Hiroshi Sato, Tsubasa Ochiai, Keisuke Kinoshita, Marc Delcroix, Tomohiro Nakatani, Shoko Araki}
\address{NTT Corporation, Japan}

\ninept

\begin{document}

\maketitle

\begin{abstract}
Target speaker extraction, which aims at extracting a target speaker's voice from a mixture of voices using audio, visual or locational clues, has received much interest. 
Recently an audio-visual target speaker extraction has been proposed that extracts target speech by using complementary audio and visual clues.
Although audio-visual target speaker extraction offers a more stable performance than single modality methods for simulated data, its adaptation towards realistic situations has not been fully explored as well as evaluations on real recorded mixtures.
One of the major issues to handle realistic situations is how to make the system robust to clue corruption because in real recordings both clues may not be equally reliable, e.g. visual clues may be affected by occlusions.
In this work, we propose a novel attention mechanism for multi-modal fusion and its training methods that enable to effectively capture the reliability of the clues and weight the more reliable ones. Our proposals improve signal to distortion ratio (SDR) by 1.0 dB over conventional fusion mechanisms on simulated data.
Moreover, we also record an audio-visual dataset of simultaneous speech with realistic visual clue corruption and show that audio-visual target speaker extraction with our proposals successfully work on real data.

\end{abstract}


\section{Introduction}
Although recent advances in machine learning techniques have greatly improved the performance of automatic speech recognition, it remains difficult to recognize speech when it overlaps other people's voices, a problem that frequently occurs in daily life.
Most approaches proposed to tackle this problem focus on blind source separation; the mixture of speech signals is separated into each of its sources. Recently deep learning-based separation approaches such as deep clustering\cite{hershey2016deep}, permutation invariant training\cite{kolbaek2017multitalker}, TasNet\cite{luo2018tasnet}, etc. have received increased attention. 

Target speaker extraction is another promising approach to deal with overlapping speech; it uses auxiliary information about the target to extract only the target speaker's voice. Hereafter, the auxiliary information is referred to as \emph{target speaker clues}.
There are several ways to realize target speaker extraction depending on the nature of the clues used. Some approaches utilize audio clues provided by enrollment utterances of the target speaker \cite{zmolikova2017speaker,wang2018deep,delcroix2018single,delcroix2019compact}. Extraction based on these audio clues has shown promising results, but the performance tends to degrade when the voice characteristics of the target and interference are similar.
Other approaches utilize video to capture the face or mouth of the target speaker as visual clues\cite{ephrat2018looking,afouras2018conversation,gabbay2018visual,wu2019time}. Visual clue-based extraction has shown robust performance for mixtures of similar voice characteristics, but it is susceptible to the occlusions caused by the speaker's hands, masks, microphones, etc. that occur in natural dialogues. 

To overcome the shortcomings of the single modality-based approaches, audio-visual target speaker extraction has been investigated recently\cite{afouras2019my,ochiai2019multimodal,gu2020multi}.
In \cite{afouras2019my}, the target speaker information extracted from audio and video clues are summed to create multi-modal embeddings, which are then used as auxiliary information to predict a time-frequency mask that extracts the target speaker out of the mixture.
Simultaneously to \cite{afouras2019my}, the multi-modal target speaker extraction method called ``audio-visual SpeakerBeam'' \cite{ochiai2019multimodal} was proposed to use source-target attention\cite{bahdanau2015neural} for fusing target speaker information extracted from audio and video clues, instead of summing them up.

Audio-visual target speaker extraction methods have been tested on simulated data and shown to offer more robust performance than single modality methods, thanks to the use of the complementary clues.
However, its application to more realistic situations has not been fully explored. 
One of the important aspects to be considered is how to deal with clue corruption.
For example, visual clues are not accessible if the speaker's hands or microphones are blocking his or her face and mouth. Audio clues recorded under noisy conditions are also not reliable.
Such corruption of clues is often present in real recordings. Besides the recordings could be corrupted by noise and reverberation that could also hinder the extraction performance.
The results in \cite{afouras2019my} indicated that existing methods could handle intermittent occlusion of visual clues to some extent when reliable audio clues were available. However, it remains unclear whether the negative effect of the unreliable visual clues on the overall performance could be well mitigated, i.e. whether the performance of the multi-modal system with clean audio clues and unreliable visual clues was not significantly lower than the performance of a system relying on the clean audio-clues only.
The modality attention mechanism was introduced in \cite{ochiai2019multimodal} with the expectation of better mitigating the effect of unreliable clues, but it was only tested on conditions where both clues were reliable and thus its effectiveness with corrupted clues has not been verified.

In this work, we focus on dealing with such a clue corruption towards more realistic applications. 
To deal with the corruption, it is desirable to utilize more informative clues while ignoring unreliable ones selectively at every time instance, according to their reliability.
We realized that the simple attention mechanism introduced in \cite{ochiai2019multimodal} was insufficient to achieve such a reliability-based clue selection due to imbalance in the norm of the contributions from the different modalities, which hindered the desirable behaviors of attention mechanisms. Therefore, we propose a novel modality attention approach that introduces a normalization mechanism to correct the norm imbalance problem of the conventional method.
Besides, we also propose two multi-task training methods that introduce a loss term on the attention-based modality fusion module to explicitly consider the reliability of the clues in the training phase, helping the attention module become more fully aware of clue reliability.

Another important aspect of this work is the evaluation of the multi-modal target speaker extraction methods on real-recordings to confirm its potential for realistic applications.
Unlike previous studies that used only simulated mixtures, we created an audio-visual dataset by recording the simultaneous speech of two speakers. 
The recordings include video of the speaker's face with natural occlusions by hand motions for evaluating the ability to deal with realistic visual-clue corruption. 
Since the amount of real recorded data is relatively small, we cannot train a system using only real recordings. Instead, we perform model adaptation to adapt a base model trained on a large simulated dataset to real recording conditions. With such a scheme, we show that audio-visual target speaker extraction can deal with real recorded mixtures.

The contribution of this work is twofold. 
\begin{enumerate}
  \item We propose a novel normalized modality attention mechanism and two multi-task training methods in order to enable the attention mechanism to select clues based on their reliability. 
  Our proposals outperform previous attention-based and summation-based fusion and maintain the performance even when either of the clues is corrupted.
  \item We conduct evaluations on real recorded mixtures with realistic occlusions and show that the audio-visual target speaker extraction performs well in real conditions. We investigate model adaptation techniques to adapt an audio-visual speech extraction model trained on a large amount of simulated data to the real recording condition for which a limited amount of data is available.
\end{enumerate}
\section{Conventional Methods\label{section:conventional}}
\subsection{General Framework}
Fig.~\ref{fig:framework} shows the general framework of the audio-visual target speaker extraction network called audio-visual SpeakerBeam proposed in \cite{ochiai2019multimodal}.
SpeakerBeam extracts the waveform signal of target speaker, $\hat{\mbox{\boldmath $x$}}_s \in \mathbb{R}^{T}$ from the mixture $\mbox{\boldmath $y$} \in \mathbb{R}^{T}$ as follows:
\begin{equation}
    \label{eq:target}
    \hat{\mbox{\boldmath $x$}}_s = {\rm SpeakerBeam}(\mbox{\boldmath $y$}, \mbox{\boldmath $c$}_s^A, \mbox{\boldmath $C$}_s^V ),
\end{equation}
where $s$ is the index of the target speaker, $T$ is the number of samples of the mixture, $\mbox{\boldmath $c$}_s^A$ corresponds to the audio clue provided by enrollment utterances of the target speaker, and
$\mbox{\boldmath $C$}_s^V$ to the visual clue derived from a video capturing the face region of the target speaker.

The network ${\rm SpeakerBeam(\cdot)}$ is composed of two parts: a \emph{target extraction network} and a \emph{clue extraction network}.
The target extraction network adopts a structure inspired by TasNet \cite{luo2019conv,delcroix2020improving} that conducts time-domain separation and has outperformed time-frequency mask-based approaches.
TasNet consists of three parts: encoder, separator and decoder. The encoder converts mixture waveform $\mbox{\boldmath $y$}$ into optimized representations $\mbox{\boldmath $Y$}$.
The separator calculates masks $\mbox{\boldmath $M$}_s$ from $\mbox{\boldmath $Y$}$ which are subsequently applied to $\mbox{\boldmath $Y$}$ by an element-wise product operation. The separator is composed of several stacked 1-D convolutional blocks followed by a 1x1 Convolution layer and a nonlinear activation function. 
The masked representations are then converted into extracted audio in raw waveform $\hat{\mbox{\boldmath $x$}}_s \in \mathbb{R}^{T}$ by the decoder.

While TasNet was developed for speech separation, here we aim at extracting only speech of the target speaker.
To guide SpeakerBeam toward extracting only the target speaker,
the processing of the network was conditioned on the \emph{audio-visual target speaker clues},
$\mbox{\boldmath $Z$}_s^{AV}$ that is computed using the ``clue extraction network'' based on audio and visual clues (see \ref{sec:av_feature_extraciton} for the details of the clue extraction network and $\mbox{\boldmath $Z$}_s^{AV}$ calculation). 
The audio-visual target speaker clues $\mbox{\boldmath $Z$}_{s}^{AV} \in \mathbb{R}^{T_e \times D_e}$ were incorporated into the TasNet architecture after $N_L$ convolutional blocks of the separator, by an element-wise product operation $\odot$ as follows\cite{delcroix2020improving}:
\begin{equation}
    \mbox{\boldmath $X$}_s' = \mbox{\boldmath $Y$}^{\prime} \odot \mbox{\boldmath $Z$}_{s}^{AV},\label{eq:element-wise-multiplication}
\end{equation}
where $\mbox{\boldmath $Y$}^{\prime} \in \mathbb{R}^{T_e \times D_e}$ is an intermediate representation of the mixture,
and $\mbox{\boldmath $X$}_s' \in \mathbb{R}^{T_e \times D_e}$ is that of the target speaker. $T_e$ and $D_e$ are the number of frames and dimension of the hidden representation respectively.
In previous studies such as \cite{delcroix2019compact}, 
it was found SpeakerBeam works effectively 
when the clue information is provided to the extraction network with
the element-wise product as in eq.~(\ref{eq:element-wise-multiplication}).

\subsection{Extraction of audio-visual clues}
\label{sec:av_feature_extraciton}
The audio-visual target speaker clues $\mbox{\boldmath $Z$}_s^{AV}$ is calculated from the audio clue $\mbox{\boldmath $c$}_s^A$, 
the visual clue $\mbox{\boldmath $C$}_s^V$, 
and the intermediate representation within the target extraction network $\mbox{\boldmath $Y$}^{\prime}$ as follows:
\begin{equation}
    \label{eq:clueext}
    \mbox{\boldmath $Z$}_s^{AV} = {\rm ClueExtraction}(\mbox{\boldmath $c$}_s^A, \mbox{\boldmath $C$}_s^V, \mbox{\boldmath $Y$}^{\prime}),
\end{equation}
where ${\rm ClueExtraction}(\cdot)$ is a neural network that consists of two sub-networks, hereafter called Cluenets, that compute embeddings associated with the audio and visual clues, and a ``modality fusion'' block that combine these embeddings into the audio-visual target speaker clues $\mbox{\boldmath $Z$}_{s}^{AV}$.

\subsubsection{Audio Cluenet}
For audio clues, the audio cluenet computes the audio embedding $\mbox{\boldmath $z$}_s^A \in \mathbb{R}^{D_{e}}$ from the raw wave audio clue $\mbox{\boldmath $c$}_s^A \in \mathbb{R}^{T_{a,s}}$ as follows:
\begin{equation}
    \label{eq:clunetA}
    \mbox{\boldmath $z$}_s^A = {\rm Average(conv(Encoder(}\mbox{\boldmath $c$}_s^A)))
\end{equation}
where $T_{a,s}$ denote the number of samples of the audio clue for speaker $s$.
The audio ${\rm Encoder}(\cdot)$ has identical structure as TasNet encoder and projects the raw-wave clue into an $D_e$ dimensional embedding space, to get two dimensional $\mbox{\boldmath $C$}_s^A \in \mathbb{R}^{T_{a,s}^{\prime} \times D_e}$ where $T_{a,s}^{\prime}$ denotes the number of frames of the hidden representation of the audio clue for speaker $s$.
$\mbox{\boldmath $C$}_s^A$ is subsequently fed into several convolutional layers (${\rm conv}(\cdot)$ in Eq. (\ref{eq:clunetA})) to extract features, and then averaged over time (${\rm Average}(\cdot)$ in Eq. (\ref{eq:clunetA})) to make time-invariant, thus one-dimensional, audio clue $\mbox{\boldmath $z$}_s^A$. 

\subsubsection{Visual Cluenet}
The visual embedding $\mbox{\boldmath $Z$}_s^V \in \mathbb{R}^{T_e \times D_e}$ is calculated with the visual cluenet from visual feature $\mbox{\boldmath $C$}_s^V \in \mathbb{R}^{T_v \times D_v}$ as follows:
\begin{equation}
    \mbox{\boldmath $Z$}_s^V = {\rm Upsample(conv(}\mbox{\boldmath $C$}_s^V))
\end{equation}
where $T_v$ and $D_v$ are the number of frames and the dimension of a visual feature.
We used pre-trained face recognition model Facenet\cite{schroff2015facenet,facenet_tk} to extract visual features $\mbox{\boldmath $C$}_s^V$ according to the procedure in \cite{ochiai2019multimodal}. They are then transformed using several convolutional layers ${\rm conv}(\cdot)$. Contrary to the audio clue, the visual clue $\mbox{\boldmath $Z$}_s^V$ is time-variant. ${\rm Upsample}(\cdot)$ was applied in order to align the number of frames of visual embeddings with $\mbox{\boldmath $Y$}^{\prime}$.  

\begin{figure}[tb]
 \begin{center}
  \includegraphics[width=.96\hsize]{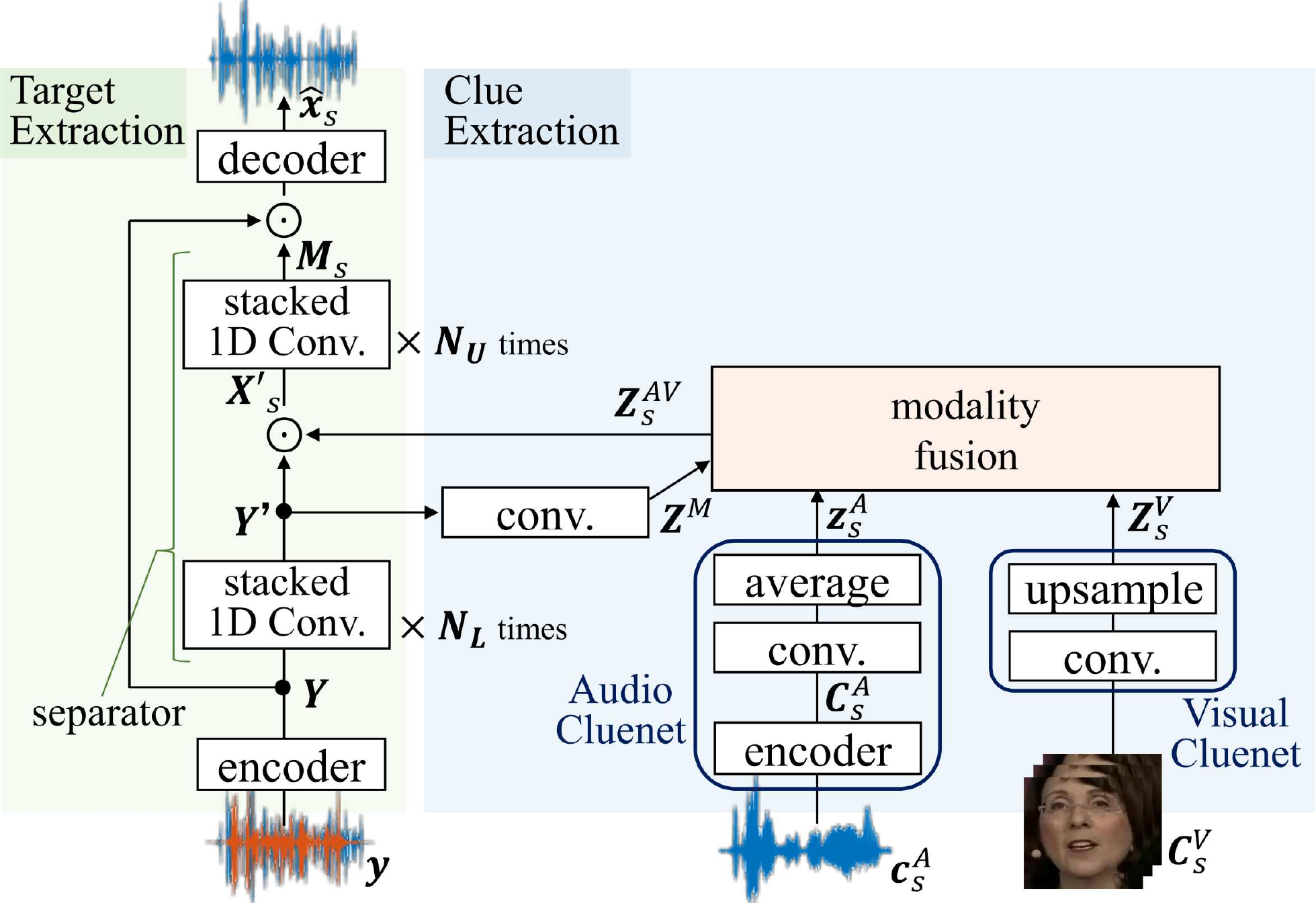}
 \end{center}
 \vspace{-18pt}
 \caption{Overview of audio-visual SpeakerBeam.}
 \vspace{-12pt}
 \label{fig:framework}
\end{figure}

\subsubsection{Modality Fusion\label{section:fusion}}
Modality fusion block incorporate each single modality embeddings $\mbox{\boldmath $z$}_{s}^{A}$ and $\mbox{\boldmath $Z$}_{s}^{V}$ into audio-visual target speaker clues $\mbox{\boldmath $Z$}_s^{AV}$ as follows:
\begin{equation}
    \mbox{\boldmath $Z$}_s^{AV} = {\rm ModalityFusion(\mbox{\boldmath $z$}_{s}^{A}, \mbox{\boldmath $Z$}_{s}^{V}, \mbox{\boldmath $Z$}_s^M)}
\end{equation}
where $\mbox{\boldmath $Z$}_s^M$ represents an embedding of the input mixture, which is calculated from an intermediate representation within the target extraction network $\mbox{\boldmath $Y$}^{\prime}$ using several convolutional layers.

Several variants have been proposed for the modality fusion block in Fig.~\ref{fig:framework}: attention fusion\cite{ochiai2019multimodal} and summation fusion\cite{afouras2019my} are examples. 
We first describe the attention fusion as it can be seen as a general case from which the summation fusion can be derived as a special case. 

\begin{figure}[tb]
 \begin{center}
  \includegraphics[width=1.00\hsize]{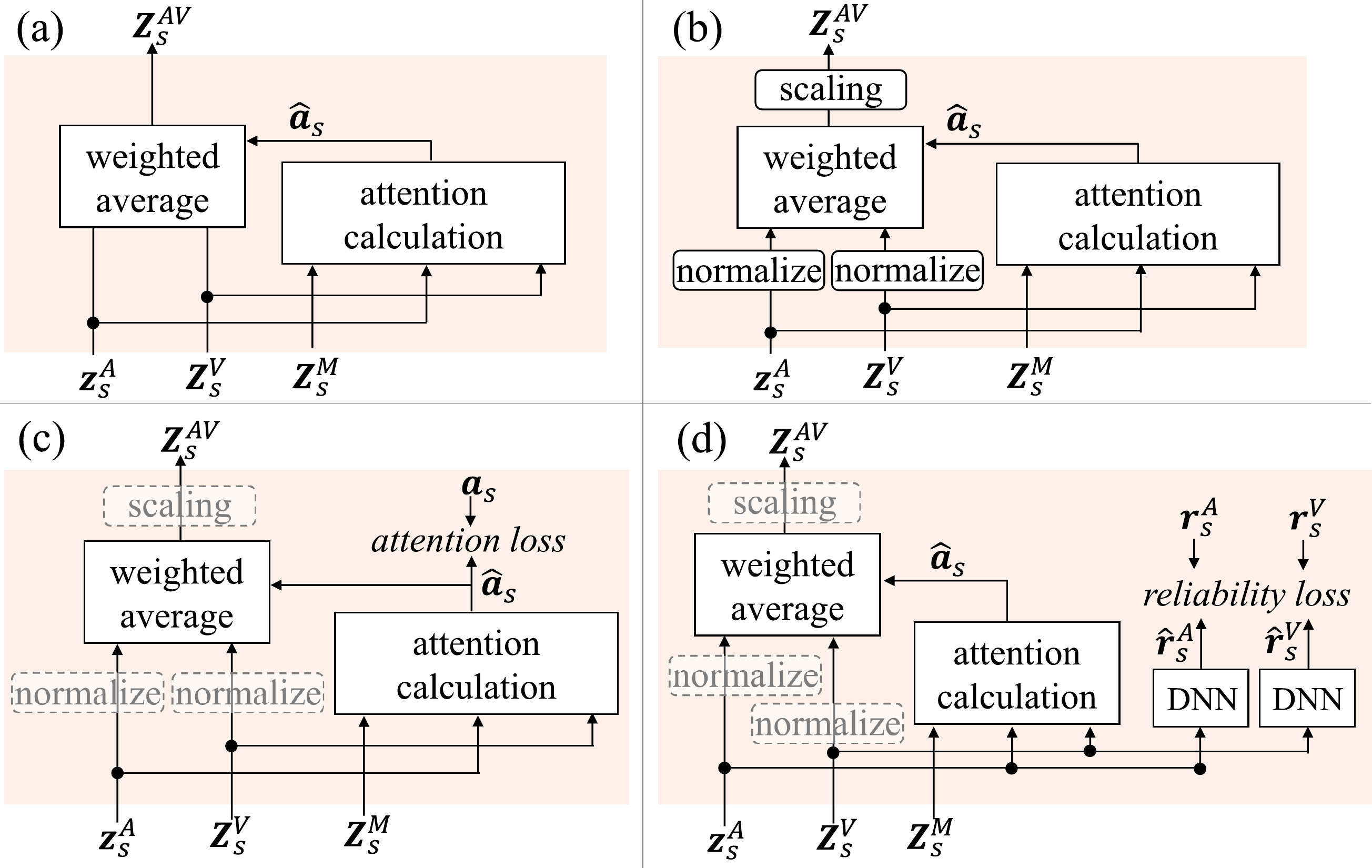}
 \end{center}
 \vspace{-20pt}
 \caption{Diagram of modality fusion methods. (a) conventional attention, (b) proposed normalized attention, (c) attention guided training and (d) clue condition aware training. System (c) and (d) can be combined with (b) normalized attention, which is described by dashed normalize and scaling blocks.}
 \vspace{-10pt}
 \label{fig:attention}
\end{figure}

Attention fusion is designed such that the network can discern the reliability of each clue and utilize a more reliable one selectively in each time instance. Fig.~\ref{fig:attention}(a) shows a conceptual diagram of the conventional attention fusion approach. 
In attention fusion, embeddings from each modality $\psi \in (A,V)$ are averaged with weight $a_{st}^{\psi}$ at each time frame $t$ as follows:
\begin{equation}
    \mbox{\boldmath $z$}_{st}^{AV} = \sum_{\psi \in (A,V)} \hat{a}_{st}^{\psi}\mbox{\boldmath $z$}_{st}^{\psi}\quad(t=1,2,\dots ,T_e)
    \label{eq:attention}
\end{equation}
where $\mbox{\boldmath $z$}_{st}^{\psi}$ is an embedding of audio and visual clues for each time frame $t\!=\!1,2,\dots,T_e$. Since the audio embedding is time-invariant, $\mbox{\boldmath $z$}_{st}^{A}$ is the same for each frame and is expressed as $\mbox{\boldmath $z$}_{st}^{A}=\mbox{\boldmath $z$}_{s}^{A}\in \mathbb{R}^{D_e}$ for each time frame. Visual embedding $\mbox{\boldmath $z$}_{st}^{V} \in \mathbb{R}^{D_e}$ is the $t$-th time frame of $\mbox{\boldmath $Z$}_{s}^{V}\!=\!\{\mbox{\boldmath $z$}_{st}^{V};t\!=\!1,2,\dots,T_e\}$. 
Following the same notation, $\mbox{\boldmath $z$}_{st}^{AV}$ is the $t$-th time frame element of $\mbox{\boldmath $Z$}_{s}^{AV}\!=\!\{\mbox{\boldmath $z$}_{st}^{AV};t\!=\!1,2,\dots,T_e\}$.
The attention value $\mbox{\boldmath $\hat{a}$}_{st} = [\hat{a}_{st}^A,\hat{a}_{st}^V] $ is determined by the ``attention calculation'' block in Fig.~\ref{fig:attention} from the audio, visual and mixture embeddings for each time frame $t$. For calculating attention value $\hat{a}_{st}^{\psi}$, the additive-attention proposed in \cite{bahdanau2015neural} was adopted. $\hat{a}_{st}^{\psi}$ is calculated as follows:
\begin{equation}
    \hat{a}_{st}^{\psi} = \frac{\exp{\epsilon e_{st}^{\psi}}}{\sum_{\psi \in (A,V)} \exp{\epsilon e_{st}^{\psi}}}
\end{equation}
\begin{equation}
    e_{st}^{\psi} = {\mathbf w} \tanh({\mathbf W}\mbox{\boldmath $z$}_{t}^{M} + {\mathbf V}\mbox{\boldmath $z$}_{st}^{\psi} + {\mathbf b})
\end{equation}
where $\mbox{\boldmath $z$}_{t}^{M} \in \mathbb{R}^{D_e}$ is the $t$-th time frame element of $\mbox{\boldmath $Z$}^{M}\!=\!\{\mbox{\boldmath $z$}_{t}^{M};t\!=\!1,2,\dots,T_e\}$.
Here ${\mathbf w}$, ${\mathbf W}$, ${\mathbf V}$ and ${\mathbf b}$ are learnable parameters, and $\epsilon$ is a sharpening factor\cite{bahdanau2015neural}.

Summation fusion proposed in \cite{afouras2019my} is another way to fuse embeddings from audio and visual modalities. It can be seen as a special case of attention fusion where the attention weights are the same for each modality regardless of the clues, i.e.  $\mbox{\boldmath $\hat{a}$}_{st} = [0.5, 0.5]$. 
In addition, the separation based on only audio or only visual clues are also special cases with $\mbox{\boldmath $\hat{a}$}_{st} = [1.0,0.0]$ and $\mbox{\boldmath $\hat{a}$}_{st} = [0.0,1.0]$, respectively.

\section{Proposed Methods\label{section:proposed}}
In this work, we focus on how to deal with clue corruption.
To this end, the attention mechanism should have the ability to select clues based on their reliability.
In order to enhance the attention mechanism acquiring this behavior, we propose 1) a normalized attention fusion method and 2) two multi-task training methods, which are described in the following sections.
\subsection{Normalized attention Fusion\label{section:norm}}
Although the attention mechanism was originally introduced to handle clue corruption by attending to more reliable clues \cite{ochiai2019multimodal}, our preliminary experiments suggested that the conventional attention mechanism explained in \ref{section:fusion} did not learn such a behavior.
As a possible cause for this issue, we found that the embeddings $\mbox{\boldmath $z$}_{st}^{\psi}$ for audio and video clues tended to take very different norms during training. In this situation, a small fluctuation in attention weights $\hat{a}_{st}$ causes a large displacement of attention output $\mbox{\boldmath $z$}_{st}^{AV}$ and hinders attention controllability, resulting in the poor extraction performance. In addition, the heavy bias in the attention values makes it difficult to interpret the attention values.

To address this issue, we propose the normalized attention mechanism shown in Fig. \ref{fig:attention}(b). Our approach is to normalize each input clue with its norm in each time frame, before integrating them as follows:
\begin{equation}
    \mbox{\boldmath $z$}_{st}^{\psi\prime} = \mbox{\boldmath $z$}_{st}^{\psi} \,/\, |\mbox{\boldmath $z$}_{st}^{\psi}|
\end{equation}
The normalized clues $\mbox{\boldmath $z$}_{st}^{\psi\prime}$ are integrated as in Eq. (\ref{eq:attention}) to get $\mbox{\boldmath $Z$}_{s}^{AV\prime}$.
$\mbox{\boldmath $Z$}_{s}^{AV\prime}$ is then multiplied by constant $l$ to reflect the original norm as follows:
\begin{equation}
    \mbox{\boldmath $Z$}_{s}^{AV} = l\mbox{\boldmath $Z$}_{s}^{AV\prime}
\end{equation}
where $l$ is calculated as $l = 1/(\sum_{\psi \in (A,V)} 1/|\mbox{\boldmath $z$}_{st}^{\psi}|)$.
In addition to resolving norm imbalance, normalized attention allows the value of inter-modal attention to provide interpretability. For example, $a_{st}^{A} = a_{st}^{V} = 0.5$ means both clues are almost equally reliable and $a_{st}^{A} >> a_{st}^{V}$ indicates that the audio clue is more reliable than the visual clue.

\subsection{Multi-task Training}
\subsubsection{Attention Guided Training\label{section:mse}}
In order to train the attention mechanism more efficiently, we also introduce multi-task loss that guides attention explicitly by directly giving the ideal attention value according to clue condition, in addition to the usual source to distortion ratio loss as shown in Fig. \ref{fig:attention}(c).
The overall loss function $\mathcal{L}$ with the attention guided training loss is expressed as follows:
\begin{equation}
    \label{eq:attention_guided}
    \mathcal{L}= \mathcal{L}_{\rm SDR}(\mbox{\boldmath $\hat{x}$}_s, \mbox{\boldmath $x$}_s) + \alpha\, \sum_{\psi \in (A,V)} \sum_{t=1}^{T_e} \frac{\mathcal{L}_{\rm MSE}(\hat{a}_{st}^{\psi}, a_{st}^{\psi})}{T_e}
\end{equation}
where $\mbox{\boldmath $x$}_s$ is the reference signal and $a_{st}^{\psi}$ is the oracle attention value. $\mathcal{L}_{\rm SDR}(\cdot)$ and $\mathcal{L}_{\rm MSE}(\cdot)$ stand for the scale invariant source to distortion ratio\cite{luo2019conv} and mean square error loss, respectively. $\alpha$ is a multi-task weight that balance the effects of each loss term. 

Since it is unclear which attention value is appropriate in every clue condition, we calculate and append the attention prediction loss only in extreme cases as follows:
1) one of the clues was completely unreliable and the other was completely clean. 2) both clues are completely clean. In the former case, we set oracle attention to $\mbox{\boldmath $\hat{a}$}_{st} = [1.0,0.0] \,{\rm or}\, [0.0,1.0]$. In the latter case we adopt $\mbox{\boldmath $\hat{a}$}_{st} = [0.5,0.5]$ to improve interpretability. When the oracle attention is not defined, we only used the first-term of Eq. (\ref{eq:attention_guided}).
Although attention guided training can be used with normalized training, this training method itself would mitigate the norm imbalance problem.

\subsubsection{Clue Condition Aware Training}
In attention guided training, oracle attention values are explicitly given, but their appropriate values are not necessarily evident. In order to make the attention module aware of clue reliability without explicitly controlling its behavior, we propose clue condition aware training as shown in Fig.~\ref{fig:attention}(d). In order for embeddings $\mbox{\boldmath $z$}_{st}^{\psi}$ to contain clue reliability information, we introduce \emph{clue condition prediction networks} that predict clue reliability $\hat{r}_{s}^{A}$ and $\hat{r}_{st}^{V}$ from each embedding. The clue condition prediction network consists of several linear layers.
The overall loss function with the clue condition aware training loss is expressed as follows:
\begin{equation}
    \mathcal{L}=\mathcal{L}_{\rm SDR}(\mbox{\boldmath $\hat{x}$}_s, \mbox{\boldmath $x$}_s) + \beta\,\left[ \mathcal{L}_{\rm MSE}(\hat{r}_{s}^{A}, r_{s}^{A}) + \sum_{t=1}^{T_e} \frac{\mathcal{L}_{\rm MSE} ( \hat{r}_{st}^{V}, r_{st}^{V})}{T_e} \right]
\end{equation}
where $r_{s}^{A}$ and $r_{st}^{V}$ represents oracle clue reliability (see \ref{section:datasets} for the definition). $\beta$ is a multi-task weight. 
Clue condition aware training complements normalized attention.

\begin{figure}[bt]
 \begin{center}
  \includegraphics[width=0.53\hsize]{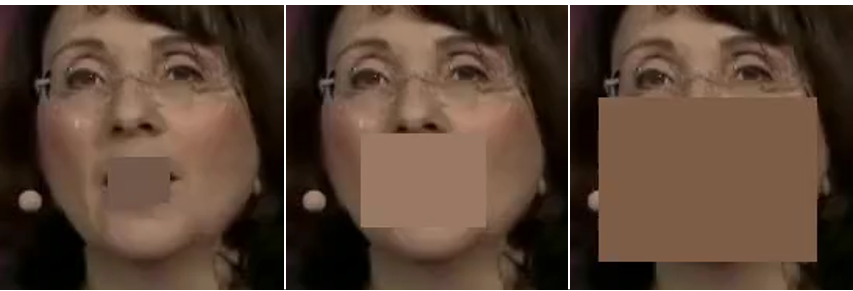}
 \end{center}
 \vspace{-22pt}
 \caption{Samples of visual clue occlusion. The photos correspond to 40x30, 80x60 and 140x105 pixel sized mask.}
 \vspace{-15pt}
 \label{fig:tedmask}
\end{figure}

\section{Experiments}

\begin{table*}[tb]
\centering
\caption{The extraction performance for the simulated dataset in SDR [dB]. System (5)-(9) are the proposals and all outperform conventional attention fusion (4) and summation fusion (3) approaches. The mask sizes of med. and full correspond to 80x60 and 188x188 pixels, respectively. ``Intermittent full'' is the condition where half of the frames are corrupted by full sized masks.}
\label{tab:ted-sdr}
\scalebox{0.92}[0.92]{
\begin{tabular}{cllc|rrrrrrrr|r}
\hline
 & \multirow{4}{*}{fusion} & \multirow{4}{*}{\begin{tabular}[c]{@{}l@{}}multitask\\ training\end{tabular}} & \multirow{4}{*}{normalize} & \multicolumn{8}{c|}{visual clues (mask size)} & \multicolumn{1}{c}{average} \\
 &  &  &  & \multicolumn{1}{c|}{no mask} & \multicolumn{2}{c|}{no mask} & \multicolumn{1}{c}{med.} & \multicolumn{1}{c|}{full} & \multicolumn{3}{c|}{intermittent full} & \multicolumn{1}{l}{} \\ \cline{5-12}
 &  &  &  & \multicolumn{8}{c|}{audio clues (SNR)} & \multicolumn{1}{l}{} \\
 &  &  &  & \multicolumn{1}{c|}{clean} & \multicolumn{1}{c}{0 dB} & \multicolumn{1}{c|}{-20 dB} & \multicolumn{2}{c|}{clean} & \multicolumn{1}{c}{clean} & \multicolumn{1}{c}{0 dB} & \multicolumn{1}{c|}{-20 dB} & \multicolumn{1}{l}{} \\ \hline
 & \multicolumn{3}{c|}{mixture} & \multicolumn{8}{c|}{0.1} & 0.1 \\ \hline
(1) & audio &  &  & \multicolumn{1}{r|}{15.1} & 14.8 & \multicolumn{1}{r|}{3.5} & 15.1 & \multicolumn{1}{r|}{15.1} & 15.1 & 14.8 & 3.5 & 12.1 \\
(2) & visual &  &  & \multicolumn{1}{r|}{15.3} & 15.3 & \multicolumn{1}{r|}{15.3} & 13.1 & \multicolumn{1}{r|}{-1.5} & 13.7 & 13.7 & 13.7 & 12.3 \\
(3) & sum &  &  & \multicolumn{1}{r|}{15.4} & 15.4 & \multicolumn{1}{r|}{15.3} & 14.8 & \multicolumn{1}{r|}{14.4} & 14.7 & 14.6 & 13.8 & 14.8 \\
(4) & attention &  &  & \multicolumn{1}{r|}{15.4} & 15.4 & \multicolumn{1}{r|}{15.3} & 14.8 & \multicolumn{1}{r|}{14.5} & 15.1 & 14.9 & 12.4 & 14.7 \\ \hline
(5) & attention & att. guided &  & \multicolumn{1}{r|}{15.9} & 15.9 & \multicolumn{1}{r|}{15.9} & 15.3 & \multicolumn{1}{r|}{14.8} & 15.6 & 15.6 & 14.4 & 15.4 \\
(6) & attention & clue cond. aware &  & \multicolumn{1}{r|}{15.9} & 15.9 & \multicolumn{1}{r|}{15.9} & 15.4 & \multicolumn{1}{r|}{14.9} & 15.7 & 15.6 & 15.1 & 15.6 \\
(7) & attention &  & \checkmark & \multicolumn{1}{r|}{\textbf{16.1}} & \textbf{16.1} & \multicolumn{1}{r|}{\textbf{16.0}} & 15.5 & \multicolumn{1}{r|}{15.1} & 15.7 & 15.5 & 15.2 & 15.6 \\
(8) & attention & att. guided & \checkmark & \multicolumn{1}{r|}{\textbf{16.1}} & \textbf{16.1} & \multicolumn{1}{r|}{\textbf{16.0}} & \textbf{15.7} & \multicolumn{1}{r|}{\textbf{15.2}} & \textbf{15.9} & \textbf{15.8} & \textbf{15.3} & \textbf{15.8} \\
(9) & attention & clue cond. aware & \checkmark & \multicolumn{1}{r|}{\textbf{16.1}} & \textbf{16.1} & \multicolumn{1}{r|}{\textbf{16.0}} & 15.6 & \multicolumn{1}{r|}{\textbf{15.2}} & 15.8 & 15.7 & 15.2 & 15.7 \\ \hline
\end{tabular}
}
\vspace{-10pt}
\end{table*}

\subsection{Datasets\label{section:datasets}}
\subsubsection{Simulated Dataset Based on LRS3-TED}
We created a simulated dataset from the Lip Reading Sentences 3 (LRS3-TED) audio-visual corpus. 
For training, we prepared a ``clean-clue set'' where both clues were unobstructed and intact and a ``corrupted-clue set'' where one of the clues was corrupted to various degrees. ``Clean-clue set'' was prepared using basically the same procedure used in \cite{ochiai2019multimodal}.
For ``corrupted-clue set'', we used the same mixture and clue pairs as the ``clean-clue set'' but randomly corrupted one of the clues. Half of the visual clue corrupted data were rendered totally unreliable by setting an artificial mask that occluded the entire region of the video, and the other half were partially unreliable with random-sized rectangular masks from 40x30 pixels to 140x105 pixels, within the 188x188 pixel-sized face region, centered on the mouth position as detected by \cite{schroff2015facenet,facenet_tk}, see Fig.~\ref{fig:tedmask}. 
Similarly, one half of the audio clue disabled data were made totally unreliable with white noise added to yield the signal to noise ratio (SNR) of -20 dB, and the other half were made partially unreliable with SNR varying from -20 dB up to 20 dB. 
In attention guided training, we calculated the multi-task loss only for the totally unreliable clues. For clue condition aware training, we define the oracle clue reliability of each visual or audio clue as $r_{st}^V=(l_{s})/(188\times4)$ where $l_{s}$ denotes the perimeter of the mask in pixels and $r_s^A=({\rm SNR}_s+20)/40$.
We conducted training with ``augmented-clue set'' that consisted of ``clean-clue set'' and ``corrupted-clue set''.

For the evaluation, we similarly prepared clean and corrupted dataset with several degrees of corruption. 
Although the visual clue occlusions were time-invariant in the training phase, i.e. same mask size for every frame, we also evaluated the performance using intermittent masks. We randomly selected multiple consecutive frames and applied masking. Half of the frames were selected and occluded by full-sized masks, so visual clues were completely unreliable half of the time.

\subsubsection{Real Recordings}
In order to verify whether the attention proposals can deal with real recorded mixtures with natural time-varying occlusions, we created an audio-visual dataset of simultaneous speech by two speakers.
The dataset was recorded under three recording conditions: 1) speakers do not occlude face with hands (\emph{without occlusion}), 2) speakers occlude their face during speech (\emph{full occlusion}) and 3) speakers move their hands freely so that the occlusion occurred intermittently (\emph{intermittent occlusion}). Eight pairs of speakers participated in the recording; four pairs had the same gender and the others had different gender. Each pair spoke 60 utterances in Japanese in total, 20 for each recording condition.

\subsection{Domain Adaptation}
Since the model trained on LRS3-TED had a large domain mismatch against the real-recordings e.g. language, lighting conditions, speaking style, etc., we conducted domain adaptation with real recordings.
In addition to the real recorded mixtures, we also recorded the same amount of non-mixed speech from eight speakers who did not attend the recordings of the real mixtures.
Based on the non-mixed speech, we generated 10,000 mixtures for the adaptation set, which contains the same three occlusion conditions as the real recorded mixtures.

As a multi-task training, we conducted attention guided training for the real dataset by assuming that both audio and visual clues were always clean in ``without occlusion'' condition and that the audio clues were always clean while the visual clues were always unreliable in ``full occlusion'' condition. 
We defined the oracle attention value at $\mbox{\boldmath $\hat{a}$}_{st}=[0.5,0.5]$ and $\mbox{\boldmath $\hat{a}$}_{st} = [1.0,0.0]$ for ``without occlusion'' and ``full occlusion'' condition, respectively. 
We did not use the multi-task loss term for ``time-variant occlusion'' condition, thus we set $\alpha = 0$ for data recorded in this condition.

\subsection{Experimental Settings}
For all experiments, the network configurations were the same except for the modality fusion block. We set hyperparameters of the target extraction network as follows: ${\rm N}=256$, ${\rm L}=20$, ${\rm B} = 256$,  ${\rm R}=4$, ${\rm X}=8$, ${\rm H}=512$ and ${\rm P} = 3$ following the notation in \cite{luo2019conv}, and $N_F=1$. 
All the cluenets had the same structure; three times of convolutional layers over the time axis (256 channels, 7x1, 5x1 and 5x1 kernel size for each layer and 1x1 stride) and layer normalization, followed by a linear layer (256 channels). 
Clue condition prediction networks consisted of three linear layers (256 channels for each hidden layer) with rectified linear unit (ReLU) activation after each hidden layer and sigmoid function after the output layer.

For training and adaptation, we adopted the Adam algorithm\cite{kingma2014adam}.
The learning rate was 5e-4 and 1e-4 for training with simulated data from scratch and adaptation with real data, respectively.
For LRS3-TED, we trained the models for 50 epochs and chose the best performing model based on a development set.
For adaptation, we conducted training for 20 epochs.
For hyperparameters, we set $\alpha=10$, $\beta=5$ to make the range of each loss term roughly the same scale, and $\epsilon=2$. 
The evaluation metric is scale invariant SDR calculated by BSSEval toolbox~\cite{vincent2006performance} and all the results were averaged over each condition of evaluation. 
Since we cannot acquire separated signals for the mixed speech of real data, we used close-talk recordings as the reference signals.
\subsection{Results and Discussion}
\subsubsection{Simulated Dataset Based on LRS3-TED\label{section:ted}}
Table~\ref{tab:ted-sdr} shows the SDR [dB] of extracted speech for simulated LRS3-TED. The result includes the conditions where both clues are clean, one of the clues is corrupted and the other is clean, and where the visual clue is intermittently occluded while the audio clue is clean or corrupted. The ``audio'' and ``visual'' in the ``fusion'' column stands for the separations with only audio clues or visual clues, and ``sum'' and ``attention'' stands for the multi-modal methods with summation and attention fusion, all respectively.
The result shows that the proposed approaches (5-9) all offered improved performance, especially the combination of normalized attention with attention guided training (8). The extraction performance of (8) was 0.96 dB better than sum (3) and 1.04 dB better than the conventional attention (4), on average. As other metrics, perceptual evaluation of speech quality (PESQ) and short-time objective intelligibility (STOI) were also calculated and improved from 2.45 and 0.939 in conventional attention (4) to 2.61 and 0.947 in the best proposed system (8) on average.
The improvement is especially large in intermittent mask conditions. The conventional systems (3-4) exhibited significant drops in performance with such time-variant visual occlusions, while the proposed systems performed robustly even with corrupted audio clues. It seems that attention proposals help to deal with clues that dynamically changes in quality, despite the model being only trained on time-invariant occlusions.

It is worth noting that the performance of the conventional methods (3-4) in the ``full mask'' condition fell behind the audio-only extraction (1) (14.4 dB and 14.5 dB v.s. 15.1 dB). 
In contrast, the best multi-modal model (8) yielded comparable SDR than the audio-only method (1) even when visual clues were totally corrupted (15.2 dB v.s. 15.1 dB). This observation suggests that our proposal can mitigate the performance degradation caused by unreliable clues. Additionally, proposal (8) significantly outperformed the visual-only method (2) even when audio clues were mostly unreliable (16.0 dB v.s.~15.3 dB). This further improvement is probably due to the multi-task effect of multi-modal training.

\begin{figure}[t]
 \begin{center}
  \includegraphics[width=1.00\hsize]{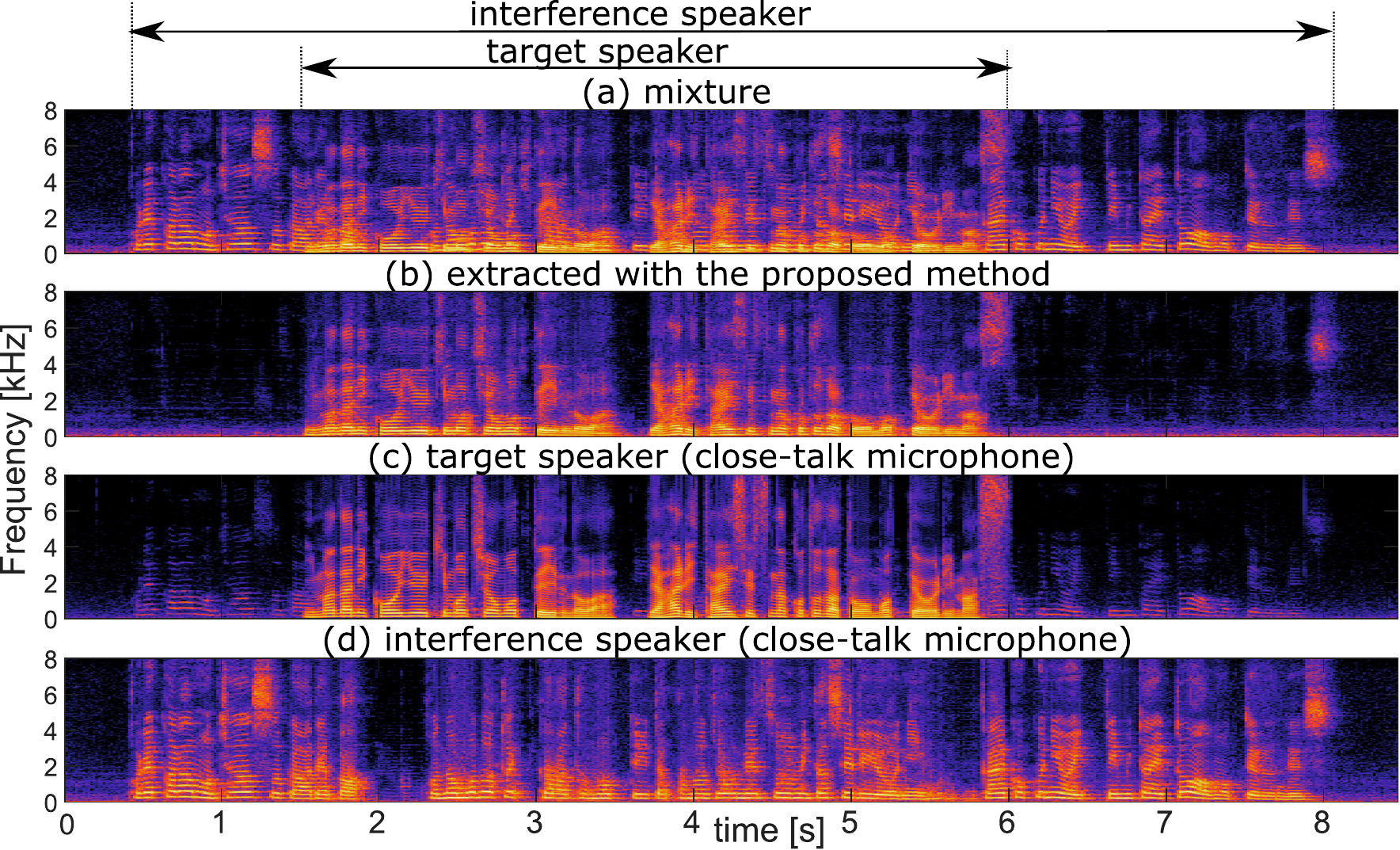}
 \end{center}
 \vspace{-22pt}
 \caption{An example of speech extracted by normalized attention model with attention guided training. The target voice is correctly extracted from the mixture.}
 \label{fig:realspec}
 \vspace{-12pt}
\end{figure}

\subsubsection{Real Recordings}
Fig.~\ref{fig:realspec} shows an example of speech extracted from a real recorded mixture. The figure shows spectrograms of (a) the mixture recorded with a distant microphone, (b) the extracted speech with the proposed method, (c) the target speaker recorded with a close-talk microphone, and (d) the interference speaker recorded with a close-talk microphone. The references (c-d) are not true target signals since it is unavailable in real recordings.
This result was obtained by the normalized attention model with attention guided training. 
From the result, we can confirm that the target voice is successfully extracted from the mixture. 
More qualitative examples are also available on the website\footnote{\url{http://www.kecl.ntt.co.jp/icl/signal/member/demo/audio_visual_speakerbeam.html}}.

Table~\ref{tab:real-sdr} shows the SDR [dB] of extracted speech. Since the ``true'' target signals are unavailable for the real mixtures, these values were calculated using close-talk microphone signals as reference signals, and are thus only indicative because the reference contains some leakage of the interference speaker as seen in Fig. \ref{fig:realspec}-(c). Evaluations on other metrics including speech recognition accuracy will be part of our future works. 

The ``audio'' and ``visual'' in the table represents the model with single modality clues, and ``attention (proposal)'' represents the normalized attention model with attention guided training; domain adaptation was conducted for each of them. The system with proposed attention extracted the target voice significantly better than the single modality systems for all occlusion conditions. This result demonstrates that audio-visual target speaker extraction with the proposed attention mechanism is also valid for real recorded mixtures even with realistic visual clue occlusions. It must also be noted that the domain adaptation improved the performance significantly for the proposed method. The result indicates that the domain adaptation could mitigate the mismatch conditions, even with a relatively limited amount of adaptation data.

\begin{figure}[t]
 \begin{center}
  \includegraphics[width=1.00\hsize]{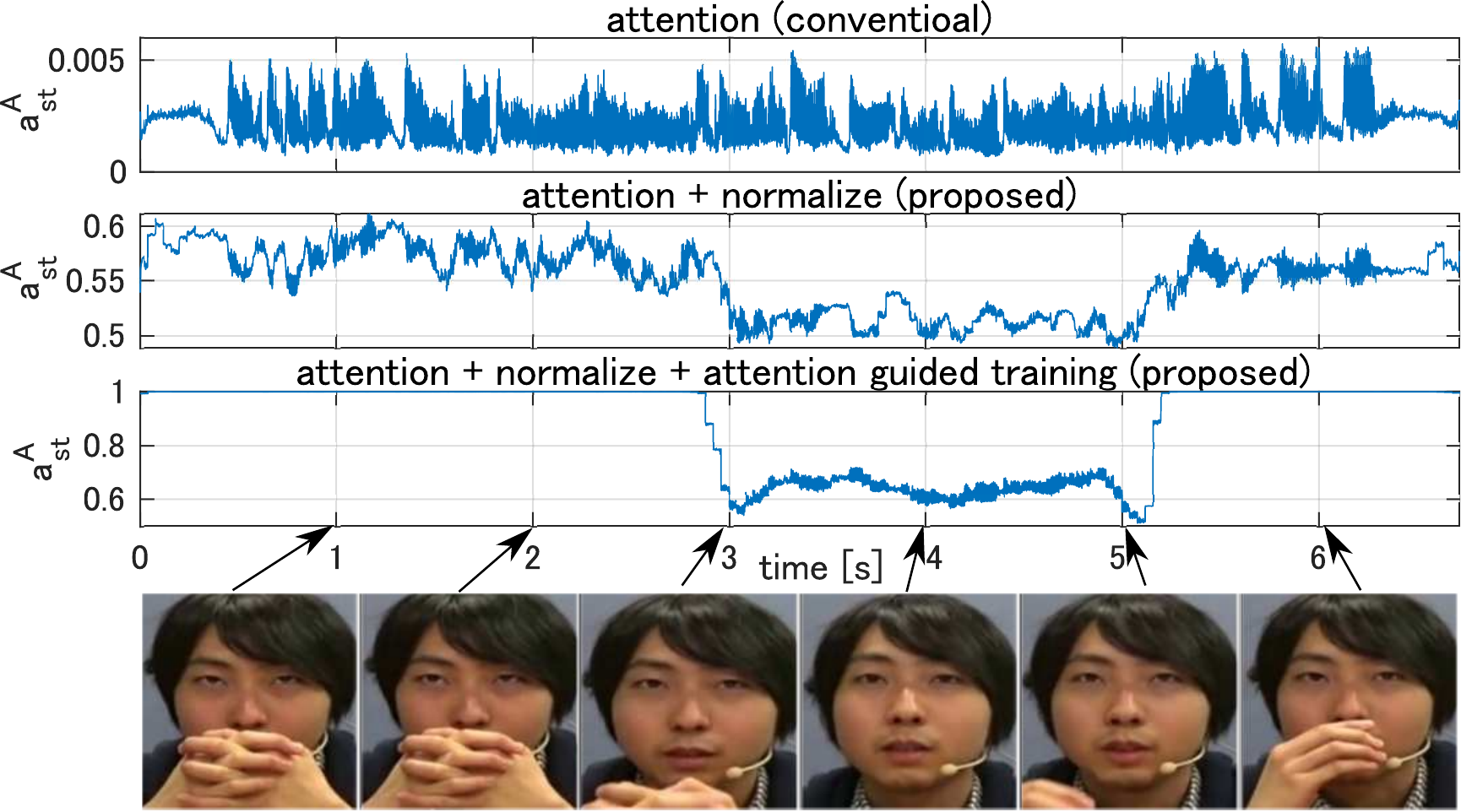}
 \end{center}
 \vspace{-21pt}
 \caption{An example of attention values to the audio modality $a_{st}^A$ . The proposed attention mechanism captured the reliability of visual clues and yielded interpretable results.}
 \vspace{-13pt}
 \label{fig:realatt}
\end{figure}

Fig.~\ref{fig:realatt} shows an example of the attention values of audio modality $a_{st}^A$ in response to the hand motions of a speaker. The results were obtained from domain adapted models.
The result shows that the conventional attention mechanism fails to capture the clue reliability. Additionally, the attention value was heavily biased to visual modality and consequently $a_{st}^A$ took values of the order of $10^{-3}$, due to the norm imbalance problem. 
By mitigating the norm imbalance problem, the proposed attention with normalization enabled to selectively weight audio clues when the mouth region of the visual clue was occluded, and to use audio and visual clues equally when not occluded.
Although the displacement of the attention values was relatively small in the normalized attention, the attention guided training further improved the attention behavior to realize the utilization of the whole range of the variability.
These results indicate that the proposed attention scheme successfully capture the reliability of the clues and learns to neglect corrupted clues even with real occlusions. This may help improve the extraction performance by mitigating the adverse effects of the corruption.
Also, this dynamic selection makes the attention value interpretable as we can judge visual clue reliability from the attention value displacements.

\begin{table}[b]
\centering
\vspace{-16pt}
\caption{Performance with real data in SDR [dB]. Our proposal extract the target voice with reasonable quality thanks to the model adaptation.}
\label{tab:real-sdr}
\scalebox{0.92}[0.92]{
\begin{tabular}{l|rrr}
\hline
system & \multicolumn{1}{l}{\begin{tabular}[c]{@{}l@{}}without\\ occlusion\end{tabular}} & \multicolumn{1}{l}{\begin{tabular}[c]{@{}l@{}}intermittent\\ occlusion\end{tabular}} & \multicolumn{1}{l}{\begin{tabular}[c]{@{}l@{}}full\\ occlusion\end{tabular}} \\ \hline
mixture & -0.4 & -0.4 & -0.3 \\ \hline
audio & 7.7 & 7.2 & 7.9 \\
visual & 3.1 & 0.6 & -1.2 \\ \hline
attention (proposal) & 8.9 & 8.3 & 8.2 \\
~~~~- w/o adaptation & 6.6 & 5.0 & 4.6 \\ \hline
\end{tabular}
}
\end{table}

\section{Conclusions}
Towards the more realistic application of audio-visual target speaker extraction, we focused on the clue corruption problem that naturally occurs in real-world recordings. In this work, we proposed a normalized attention mechanism and two multi-task training methods that efficiently train the attention mechanism to be aware of clue reliability. Experiments on a simulated dataset showed that both proposed methods improve the extraction performance over conventional attention fusion and summation fusion by capturing clue reliability.

We also conducted experiments on real recorded mixtures with natural visual occlusions. The results showed that our proposals with domain adaptations successfully extracted target voices from mixtures, at the same time of correctly dealing with the visual occlusions to selectively use more informative clues for each time instance.

As future works, we will carry a deeper investigation using real datasets with more training and test speakers as well as adopting other evaluation metrics such as speech recognition accuracy.
\bibliographystyle{IEEEbib}
\bibliography{mybib}
\end{document}